\documentclass[12pt,a4paper,english,superscriptaddress,aps,nofootinbib]{revtex4}
\usepackage[utf8]{inputenc}
\usepackage[T1]{fontenc}
\usepackage{amsmath,amssymb,graphicx}
\makeatletter
\usepackage{babel}
\usepackage[active]{srcltx}
\usepackage{graphicx,color}
\usepackage{changebar}
\usepackage{hyperref}
\usepackage[T1]{fontenc}
\usepackage{esint}
\usepackage{multirow}
\usepackage{dsfont}
\usepackage{ae}
\usepackage{amsmath}
\usepackage{braket}
\usepackage{mathtools}
\usepackage{slashed}
\usepackage{empheq}
\usepackage{multirow}
\usepackage{graphicx}
\usepackage{amsfonts}
\usepackage{amsmath}
\newcommand{\be}{\begin{equation}}
	\newcommand{\ee}{\end{equation}}
\newcommand{\bea}{\begin{eqnarray}}
	\newcommand{\eea}{\end{eqnarray}}

\begin{document}
	\title{Non-Hermitian fermions with effective mass}
	
\author{F. C. E. Lima}
\email{cleiton.estevao@fisica.ufc.br}
\affiliation{Universidade Federal do Cear\'{a} (UFC), Departamento do F\'{i}sica - Campus do Pici, Fortaleza, CE, C. P. 6030, 60455-760, Brazil.}

\author{L. N. Monteiro}
\affiliation{Universidade Federal do Cear\'{a} (UFC), Departamento do F\'{i}sica - Campus do Pici, Fortaleza, CE, C. P. 6030, 60455-760, Brazil.}
 
\author{C. A. S. Almeida}
\email{carlos@fisica.ufc.br}
\affiliation{Universidade Federal do Cear\'{a} (UFC), Departamento do F\'{i}sica - Campus do Pici, Fortaleza, CE, C. P. 6030, 60455-760, Brazil.}

\begin{abstract}
In this work, we readdress the Dirac equation in the position-dependent mass (PDM) scenario. Here, one investigates the quantum dynamics of non-Hermitian fermionic particles with effective mass assuming a $(1+1)$-dimension flat spacetime. In seeking a Schr\"{o}dinger-like theory with $\mathcal{PT}$ symmetry is appropriate to assume a complex potential. This imaginary interaction produces an effective potential purely dependent on mass distribution. Furthermore, we study the non-relativistic limit by adopting the Foldy-Wouthuysen transformation. As a result, that limit leads to an ordering equivalent to the ordering that describes abrupt heterojunctions. Posteriorly, particular cases of mass distribution also were analyzed. Thus, interesting results arise for the case of a linear PDM, which produces an effective harmonic oscillator and induces the emergence of bound states. For a hyperbolic PDM, an effective potential barrier emerges. However, in this case, the fermions behave as free particles with positive-defined energy.
\end{abstract}

\maketitle

\thispagestyle{empty}

\section{Introduction}

Problems of position-dependent effective mass help us to understand impurity in crystals \cite{Luttinger, Wannier, Slater}, and heterojunctions in semiconductors \cite{Bastard, Weisbuch}. Furthermore, the idea of the position-dependent effective mass (commonly called position-dependent mass or PDM) became an attractive topic for several researchers, see Refs. \cite{Mustafa0, Costa, Zare, Ghafourian, Pourali}.

The PDM concept in a non-relativistic regime is ambiguous. Indeed, this is because the momentum operator is not well defined. Therefore, it needs to symmetrize its kinetic energy operator (KEO) to study PDM. A few symmetrized KEO operator proposals are the orderings of BenDaniel and Duke \cite{BenDaniel}, Li and Kuhn \cite{LiKuhn}, Zhu and Kroemer \cite{ZhuKroemer}, Gora and Williams \cite{GoraWilliams}. These orderings previously mentioned are compressed into the equation
\begin{align}\label{vonRoos}
    \hat{\mathcal{K}}=\frac{1}{4}[m^{\alpha}(\vec{r})\hat{p}\,m^{\beta}(\vec{r})\hat{p}\,m^{\gamma}(\vec{r})+m^{\gamma}(\vec{r})\hat{p}\,m^{\beta}(\vec{r})\hat{p}\,m^{\alpha}(\vec{r})].
\end{align}
This general ordering was proposed initially by von Roos. In his proposal, the parameters $\alpha$, $\beta$, and $\gamma$ must respect the condition $\alpha+\beta+\gamma=-1$ \cite{RefVRoos}. Furthermore, it is essential to highlight that there is no unanimity on the appropriate form of the KEO (or the values of the Hermiticity parameters). Table \ref{tab1} shows the main orderings proposed in the literature.
\begin{table}[ht!]
    \centering
    \caption{Values of the hermiticity parameters.}
    \resizebox{10cm}{!}{%
    \begin{tabular}{ccc}\hline \hline
    Authors (year) & KEO parameters \\ \hline \hline
     BenDaniel and Duke (1966)   & $\alpha=\gamma=0$ and $\beta=-1$ \\
     Gora and Williams (1969)   & $\beta=\gamma=0$ and $\alpha=-1$ \\
     Zhu and Kroemer (1983) & $\alpha=\gamma=-\frac{1}{2}$ and $\beta=0$ \\
     Li and Kuhn (1993) & $\beta=\gamma=-\frac{1}{2}$ and $\alpha=0$ \\ \hline
    \end{tabular}}
    \label{tab1}
\end{table}

The KEO ambiguity only arises in non-relativistic theory. In relativistic quantum mechanics, this problem does not appear. Indeed, this is a consequence of the mass term not being coupled to the quadri-momentum operator, as shown first by one of the author of the present work \cite{Cavalcante}. It is essential to highlight that, in principle, we can apply the PDM concept in relativistic quantum mechanics \cite{Cavalcante} and non-relativistic one \cite{Dutra, DHA}. As a matter of fact, in 2000, Dutra and Almeida \cite{Dutra}, have applied the PDM concept to discuss the exact solvability of the Schr\"{o}dinger equation. It is found that even in the presence of KEO ambiguity some class of potential are exactly solved. On the other hand, the PDM arises in investigations of exact solutions of some kind of exponential mass distribution \cite{Hassanabadi}, semi-confined harmonic oscillator \cite{Quesne}, quantum gravitational effect \cite{Lawson}, and quantum information entropies \cite{Lima1}.

Generally, we expect a quantum theory to be a Hermitian theory \cite{Lima}. However, the non-Hermitian Hamiltonians are essential in studies of open quantum systems in nuclear physics or quantum optics (see Refs. \cite{R1,R2}). These non-Hermitian Hamiltonians are considered an effective subsystem within a projective subspace of the total system, which obeys conventional quantum mechanics with a Hermitian Hamiltonian \cite{R1,R2,R3}. In 1998, Bender et al. \cite{Bender, R4} showed that the theory is unitary with real energy eigenvalues obtained when a weak condition on the Hamiltonian is assumed. This condition is parity-time symmetric Hamiltonians, along with a deformed inner product \cite{Bender,R4}. Briefly, Bender et al. \cite{Bender, Bender1} showed that for a non-Hermitian quantum theory to be accepted physically, the Hamiltonian must be invariant under $\mathcal{PT}$ symmetry \cite{Bender2, Weigert}. 

Currently, several works discussing non-Hermitian models have emerged in the literature. Between these works, we can mention investigations of topology \cite{Bergholtz}, skin effect \cite{LiLee,Zhang}, quantum information and thermodynamic properties \cite{Lima}, and phase transitions in quasi-crystals \cite{Weidemann}. The list of applications related to non-Hermitian models is extensive, which motivates us to study a non-Hermitian theory for a PDM context.

Therefore, we seek to formulate a non-Hermitian theory of effective mass. To achieve our purpose it was built a non-Hermitian approach for fermions with arbitrary PDM. To exemplify the method, we choose mass profiles in which the Hamiltonian is kept invariant under the $\mathcal{PT}$ transformation and studied the quantum eigenstates of a linear PDM. In this case, it is possible to perceive that linear mass produces an effective harmonic oscillator. Moreover, for a PDM with a hyperbolic profile potential an effective barrier will arise.

We organized our paper as follows: In Sec. II it is introduced a non-Hermitian theory of effective mass inside Dirac's equation. Furthermore, the non-relativistic limit of the kinetic term is analyzed. In this limit, the kinetic contribution of the Hamiltonian is similar to the ordering proposed by Li and Kuhn \cite{LiKuhn}, as predicted by Cavalcante et al. \cite{Cavalcante}. In Sec. III, particular cases of mass dependence are studied, namely, linear mass distribution and a hyperbolic distribution. Finally, we present our findings in Sec. IV.


\section{The non-Hermitian theory for a PDM}
\label{section-2}

An active issue of condensed matter physics is the symmetrization problem of the kinetic energy operator in effective mass theory \cite{Lima1}. That is because, in this scenario, the use of the concept of effective mass can describe defects or impurities in crystals \cite{Luttinger, Wannier, Slater}. In this work, we propose an investigation of a PDM in a non-Hermitian Dirac theory.

To reach our purpose and to bypass the ambiguity of KEO, allow us to assume a relativistic system in $(1+1)$D. In this case, Dirac's equation is written as
\begin{equation}\label{eq1}
    \left[\gamma^\mu (p_\mu-eA_\mu) - m(x) \right]\psi(x,t) = 0 \, \, \, \, \, \, \text{with} \, \, \, \, \, \, \mu=0,1.
\end{equation}
Here, $p_\mu$ is the momentum, $m(x)$ is the PDM, $\gamma^\mu$ are the Dirac matrices. Furthermore, the signature metric used is $g_{\mu\nu}=(+,-)$. Besides, $A_\mu=(V(x),\vec{A}(x))$ where $V(x)$ is the electrical interaction, and $\vec{A}(x)$ is the vector potential (see Ref. \cite{Greiner}). Let us assume a system only with electrical interaction. In this case, we have $A_\mu=(V(x),0)$. Using these definitions Eq. (\ref{eq1}) is rewritten as
\begin{align}\label{eq2}
 i \mathcal{I}_2\frac{\partial \psi}{\partial t} = \left( -i \gamma^0\gamma^1 \partial_x + \gamma^0 m(x) +\mathcal{I}_2V(x)\right) \psi(x,t). 
\end{align}
To obtain Eq. (\ref{eq2}) it was necessary to use the identity operator $(\gamma^0)^{2}=-(\gamma^1)^2$. Here, we define $\mathcal{I}_2$ as
\begin{align}
    \mathcal{I}_2=\begin{pmatrix}
       1 & 0\\
       0 & 1
    \end{pmatrix}.
\end{align}
i. e., $\mathcal{I}_2$ represents a second rank identity matrix.

Eq. (\ref{eq2}) informs us that the Hamiltonian operator is
\begin{align}
    \mathcal{H}=\alpha p_x + \beta m(x)+\mathcal{I}_2V(x), \label{eq:Hamiltonian}
\end{align}
where the matrices $\alpha=\gamma^0\gamma^1$ and $\beta=\gamma^0$. From now on, let us assume a stationary theory. In this case, the Hamiltonian is not explicitly time-dependent. Therefore, it is assumed that the wave function has the form $\psi(x,t) = e^{\frac{iE}{\hbar}t} \varphi(x)$. Furthermore, choosing the Dirac matrix representation in $(1+1)$D, namely,
\begin{align}
    \gamma^0 = \begin{pmatrix} 0 & 1 \\ 1 & 0
    \end{pmatrix} \hspace{1cm} \text{and}  \hspace{1cm} \gamma^1 =\begin{pmatrix} 0 & -1 \\ 1 & 0 \end{pmatrix} \label{eq:representation}
\end{align}
we arrived at
\begin{equation}\label{HH}
    \mathcal{H}\varphi = E\varphi,
\end{equation}
with
\begin{align}
    \varphi=\begin{pmatrix}
       \varphi_1\\\varphi_2
    \end{pmatrix}.
\end{align}

In matrix form, Dirac's equation for a PDM with interaction is
\begin{align}
    \begin{pmatrix}
   -i \partial_x+V(x) & m(x) \\
   m(x) & i \partial_x+V(x)
   \end{pmatrix} \begin{pmatrix} \varphi_1 \\ \varphi_2 \end{pmatrix}
    = E \begin{pmatrix} \varphi_1 \\ \varphi_2 \end{pmatrix}.
   \label{eq42}
\end{align}

Notice that Eq. (\ref{eq42}) produces two coupled differential equations, i. e.,
\begin{align}
    - i\frac{d\varphi_1}{dx}+m(x)\varphi_2=&(E-V(x))\varphi_1,
\end{align}
and
\begin{align}
     i\frac{d\varphi_2}{dx}+m(x)\varphi_1=&(E-V(x))\varphi_2.
\end{align}
 
By algebraic manipulations, we can decoupled the $\varphi_1$ (or $\varphi_2$) component. In this case, the decoupled equation for the $\varphi_1$ component will be
\begin{align}\label{Ephi}\nonumber
    -\varphi''_1(x)+\frac{m'(x)}{m(x)}\varphi'_1(x)+\bigg[2EV(x)-V(x)^2-iV'(x)-i\frac{m'(x)}{m(x)}(E-V(x))\bigg]\varphi_1(x)=&\\
    (E^2-m(x)^2)\varphi_1(x).&
\end{align}
The prime notation represents the derivatives concerning the variable $x$. Eq. (\ref{Ephi}) was previously discussed by Jia and Dutra \cite{Sheng} to address a PDM problem with violation of $\mathcal{PT}$ symmetry.

To treat Eq. (\ref{Ephi}), allow us to consider the following dependent variable transformation, i. e.,
\begin{align}\label{Transf0}
       \varphi_1(x)=m(x)^{a}\phi_1(x).
       \end{align}

The transformation (\ref{Transf0}), leads us to
\begin{align} \label{EqTransf}\nonumber 
    -\phi_1''(x)+\frac{m'(x)}{m(x)}(1-2a)\phi'_1(x)+\bigg[a\bigg(\frac{m'(x)}{m(x)}\bigg)^2-a\frac{m''(x)}{m(x)}-a(a-1)\frac{m'(x)^2}{m(x)^2}+2EV(x)+&\\
    -V(x)^2-iV'(x)-i\frac{m'(x)}{m(x)}(E-V(x))\bigg]\phi_1(x)=(E^2-m(x)^2)\phi_1(x).&
\end{align}

Considering $a=1/2$, we arrive at
\begin{align} \label{EqTranf1}\nonumber
    -\phi''_1(x)+\bigg[m(x)^2+\frac{3}{4}\frac{m'(x)^2}{m(x)^2}-\frac{1}{2}\frac{m''(x)}{m(x)}+\bigg(2V(x)-i\frac{m'(x)}{m(x)}\bigg)E-V(x)^2-iV'(x)+&\\
    +i\frac{m'(x)}{m(x)}V(x)\bigg]\phi_1(x)=E^2\phi_1(x).&
\end{align} 

Here, allow us to define the effective potential as
\begin{align}\label{PoEff} \nonumber
    V_{\text{eff}}(E, m(x);x)=m(x)^2+\frac{3}{4}\frac{m'(x)^2}{m(x)^2}-\frac{1}{2}\frac{m''(x)}{m(x)}+\bigg(2V(x)-i\frac{m'(x)}{m(x)}\bigg)E-V(x)^2-iV'(x)+&\\
    +i\frac{m'(x)}{m(x)}V(x).&
\end{align}
Physically this effective potential is produced due to the effective behavior of the particle. Indeed, this is because the particle (or defect) behaves like a system with PDM, which generates an effective interaction. Seeking to investigate a model where effective interaction is purely dependent on mass profile, let us propose a potential
\begin{align}\label{Potential}
    V(x)=i\frac{m'(x)}{2m(x)}.
\end{align}
This choice of potential leads us to the Hermitian effective theory since the mass profile must always be positive-defined. Furthermore, complex effective potential suggests that our model has an apparent paradox, i. e., it is a non-Hermitian theory. This apparent result leads us to the hypothesis of complex energy eigenvalues. Nonetheless, as discussed by Ramos et al. \cite{Ramos}, although the system may be non-Hermitian, the eigenenergies of the system may be real. For more details on non-Hermitian theories, see Refs. \cite{Lima,Bender,Bender1,Ramos}. Indeed, this is because the effective theory given by Eq. (\ref{EqTranf1}) is a Schr\"{o}dinger-type theory that is invariant under $\mathcal{PT}$ symmetry. Besides, one notes that our Hamiltonian (\ref{eq:Hamiltonian}) has the form $\mathcal{H}\equiv \mathcal{K}+if(x)$ ($\mathcal{K}$ is the kinetic energy operator and $f(x)$ a position function). Indeed, Bender studied this class of Hamiltonian in Ref. \cite{B}. As discussed in Ref. \cite{B}, we conclude that our system guarantees common states of $\mathcal{H}$ and $\mathcal{PT}$, i. e., $\mathcal{H}\varphi=E\varphi$ and $\mathcal{PT}\varphi=\lambda\varphi$, which gives us $E=E^*$ since $[\mathcal{PT},\mathcal{H}]=0$. More details about this proof are in Ref. \cite{B}.

Finally, note that the interaction (\ref{Potential}) leads us to an effective potential of the form
\begin{align}\label{PotVEff}
    V_{\text{eff}}(m(x);\, x)=m(x)^2.
\end{align}
This effective potential is dependent purely on the mass distribution of the particle. Here it is necessary to highlight that the mass must be a real function and positive-defined. Furthermore, it is essential to mention that the effective potential profile will define whether we will have bound or free quantum states. In principle, the choice of mass profile is arbitrary. However, if $m(x)$ is such effective potential assumes a confining configuration (i. e., a potential well), the fermions will have bound states. Otherwise, only free states will exist. In the next section, we will present an example for each case.

Adopting the effective potential (\ref{PotVEff}), one obtains that the $\phi_1(x)$ component of the fermion with effective mass is described by
\begin{align}\label{DiracPDM}
    -\phi''_1(x)+m(x)^2\phi_1(x)=E^2\phi_1(x).
\end{align}

\subsection{The non-relativistic limit}

In Sec. \ref{section-2}  we developed the relativistic theory. Nonetheless, the relativistic theory and non-relativistic must be compatible \cite{Greiner}. In Ref. \cite{Cavalcante}, in the non-relativistic limit, the fermionic particle with PDM is described by the ordering of Li and Kuhn \cite{LiKuhn}. In our study, the issue arises: in the non-relativistic regime, the Hamiltonian (\ref{eq:Hamiltonian}) is equivalent to which theory of effective mass? Naturally, this questioning motivates us to seek a correspondence of our model in the non-relativistic limit. Therefore, to analyze the low-energy limit, i. e., when the kinetic energy is small compared to the rest energy, the term $\beta m(x)$ is dominant. This consideration is a consequence of speeds being very small compared to the speed of light.

To obtain the non-relativistic limit, let us start by considering
\begin{align}\label{H1}
    \mathcal{H}\to \beta m(x),
\end{align}
when $c\to\infty$. Here, $c$ is the speed of light.

Using the approach proposed by Fouldy and Wouthuysen \cite{Transf-FW}, it is possible to investigate a corresponding theory in the non-relativistic limit. Thereby, in search of a correspondence between the relativistic and non-relativistic theories, allow us to assume the Fouldy-Wouthuysen transformation in the $\phi$ spinor, namely,
\begin{align}
    \varphi(x) \rightarrow \text{e}^{iS}   \varphi(x),
\end{align}
so that the transformed Hamiltonian is
\begin{align}\label{eq:H-transformado}
    \mathcal{H}' = \text{e}^{iS}\, \mathcal{H}\,\text{e}^{-iS}.
\end{align}

Following the proposal of Foldy and Wouthuysen \cite{Transf-FW} and considering that the mass and momentum will not commute, the operator $S$ is
\begin{align}\label{S1}
    S = -\frac{i}{2} \frac{1}{\sqrt{m(x)}} \beta \alpha p_x \frac{1}{\sqrt{m(x)}}.
\end{align}
As discussed by Fouldy and Wouthuysen \cite{Transf-FW} for a constant mass and later by Cavalcante \textit{et al.} \cite{Cavalcante} for a PDM, there are no restrictions on the unitarity of the operator $S$. For more details, see Refs. \cite{Cavalcante,Transf-FW}.

Using the Baker-Hausdorff lemma \cite{Sakurai} to expand the equation (\ref{eq:H-transformado}), let us write the transformed Hamiltonian as
\begin{align}
    \mathcal{H}'=\mathcal{H}+i[S,\,\mathcal{H}]-\frac{1}{2}[S,\,[S,\,\mathcal{H}]]+\dots
\end{align}

Assuming the representation (\ref{eq:representation}), we arrive at the conclusion that
\begin{align}\label{o1}
    S=-\frac{1}{2} \begin{pmatrix}   0 & -\frac{1}{m(x)} \frac{d}{dx} + \frac{m'}{2m^2}  \\ \frac{1}{m(x)} \frac{d}{dx} - \frac{m'}{2m^2}  & 0 \end{pmatrix},
\end{align}
and
\begin{align}\label{o2}
    [S,\,\mathcal{H}] = \begin{pmatrix} \frac{d}{dx} & 0 \\ 0 & - \frac{d}{dx}    \end{pmatrix}.
\end{align}

Furthermore, one obtains
 \begin{align}\label{o3}
     -\frac{1}{2} [S,\,[S,\,\mathcal{H}]] &= \Big( \frac{1}{2m} \frac{d^2}{dx^2} - \frac{m'}{2m^2} \frac{d}{dx} - \frac{m''}{8m^2} \Big) \beta \nonumber \\
     &= -\frac{1}{4} \Big( \frac{1}{\sqrt{m(x)}}p_x\frac{1}{\sqrt{m(x)}}p_x + p_x \frac{1}{\sqrt{m(x)}} p_x \frac{1}{\sqrt{m(x)}}   \Big) \beta.
 \end{align}
To calculate the above commutators, the order terms $\mathcal{O}(m^{-3})$ or higher are negligible.

Adopting the operators $S$ (\ref{S1}),  $\mathcal{H}$ (\ref{H1}) and their commutation relations (Eqs. (\ref{o1}), (\ref{o2}), and (\ref{o3})), one arrives at
\begin{align}
    \mathcal{H}'&=\mathcal{H}+i[S,\,\mathcal{H}]-\frac{1}{2} [S,\,[S,\,\mathcal{H}]] \nonumber \\
    & = \Big( \alpha p_x + \beta m(x)  \Big) - \alpha p_x - \frac{1}{4} \Big( \frac{1}{\sqrt{m(x)}}p_x\frac{1}{\sqrt{m(x)}}p_x + p_x \frac{1}{\sqrt{m(x)}} p_x \frac{1}{\sqrt{m(x)}}   \Big) \beta  \nonumber \\
    & = \Big[m(x) - \frac{1}{4} \Big( \frac{1}{\sqrt{m(x)}}p_x\frac{1}{\sqrt{m(x)}}p_x + p_x \frac{1}{\sqrt{m(x)}} p_x \frac{1}{\sqrt{m(x)}} \Big) \Big]\beta. 
\end{align}
Perceive that this result is the explicit contribution of rest energy added to the kinetic energy. So, we write the non-relativistic KEO as
\begin{align}
    \hat{\mathcal{K}}=\frac{1}{4} \Big( \frac{1}{\sqrt{m(x)}}p_x\frac{1}{\sqrt{m(x)}}p_x + p_x \frac{1}{\sqrt{m(x)}} p_x \frac{1}{\sqrt{m(x)}} \Big).
\end{align}
Therefore, the non-relativistic Hamiltonian will be
\begin{align}\label{HH1}
    \mathcal{H}=\frac{1}{4} \Big( \frac{1}{\sqrt{m(x)}}p_x\frac{1}{\sqrt{m(x)}}p_x + p_x \frac{1}{\sqrt{m(x)}} p_x \frac{1}{\sqrt{m(x)}} \Big)+V(x).
\end{align}
That is the Hamiltonian ordered by Li and Kuhn \cite{LiKuhn} for a PDM. Looking at Tab. \ref{tab1}, the Hermiticity parameters are $\beta=\gamma=-\frac{1}{2}$ and $\alpha=0$. Indeed, in Ref. \cite{Cavalcante}, the authors discuss the ordering of Li and Kuhn. They consider a quantum system with a dimension higher in another fermionic representation and obtain the same result. Furthermore, the profile of KEO (and the Hamiltonian) in the non-relativistic limit is extremely relevant. Basically, this is because the electron transmission phenomenon in semiconductor heterostructures is sensitive to the Hermicity parameters \cite{Cavalcante}. Therefore, our results also correspond to the theory proposed by Li and Kuhn \cite{LiKuhn} for semiconductor heterostructures.


\section{Some particular cases of PDM}

Throughout this section, we will consider two particular mass profiles, i. e., a linear mass and a hyperbolic. Indeed, these mass profiles are convenient because they describe systems of interest in condensed matter physics and solid-state physics. That is because the linear mass can describe, for example, the electrons in a graphene sheet. This description is possible because the effective mass of these systems is linear, and its potential is harmonic-like. For more details on linear PDM, see Ref. \cite{Gomes}. The second type of effective mass we chose is hyperbolic mass. Particularly, the hyperbolic mass profile considered is the mass of a soliton. Thus, our system will describe, for example, a defect that propagates in a crystal lattice (solid-state physics system), keeping its shape and energy unchanged  \cite{Gomes, Rajaraman}. Motivated by these applications, we will now understand the quantum dynamics of non-Hermitian fermions with this PDM.

\subsection{Linear mass generating an harmonic effective potential}

\begin{figure}[ht!]
    \centering
    \includegraphics[height=6.5cm,width=8cm]{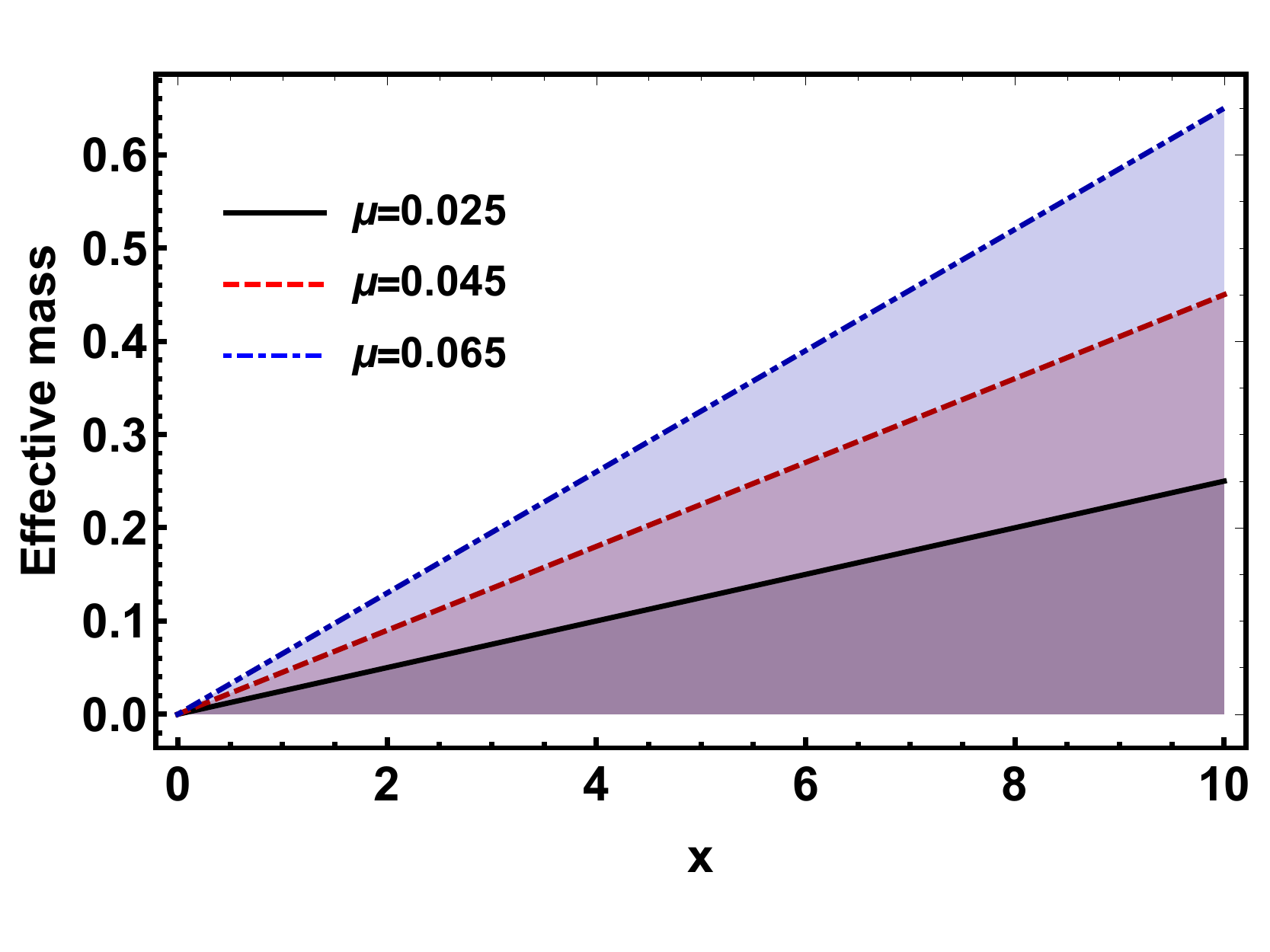}
    \vspace{-1.1cm}
    \caption{Linear PDM with several values of $\mu$.}
    \label{Lfig1}
\end{figure}

\begin{figure}[ht!]
    \centering
    \includegraphics[height=6.5cm,width=8cm]{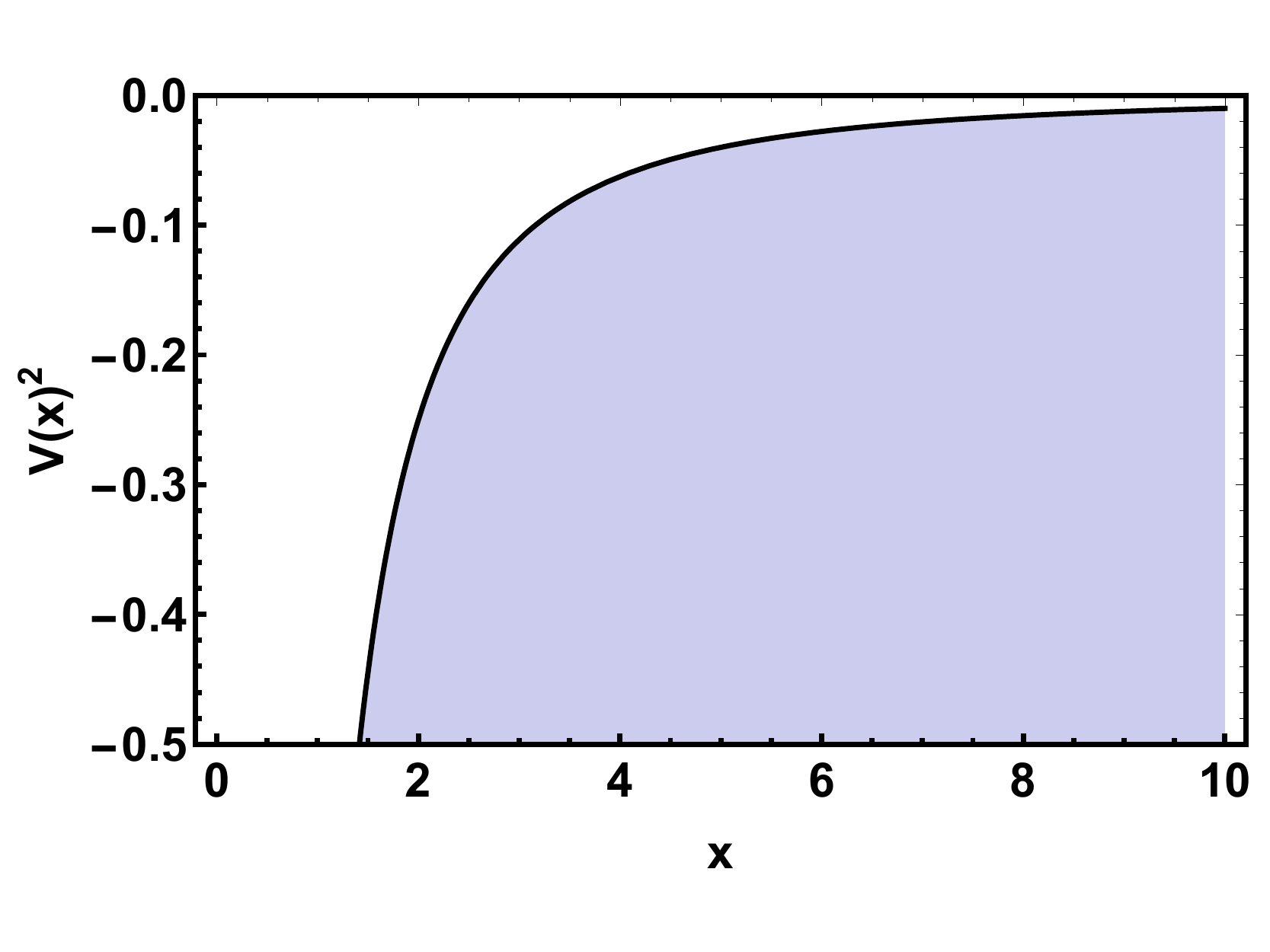}
    \includegraphics[height=6.5cm,width=8cm]{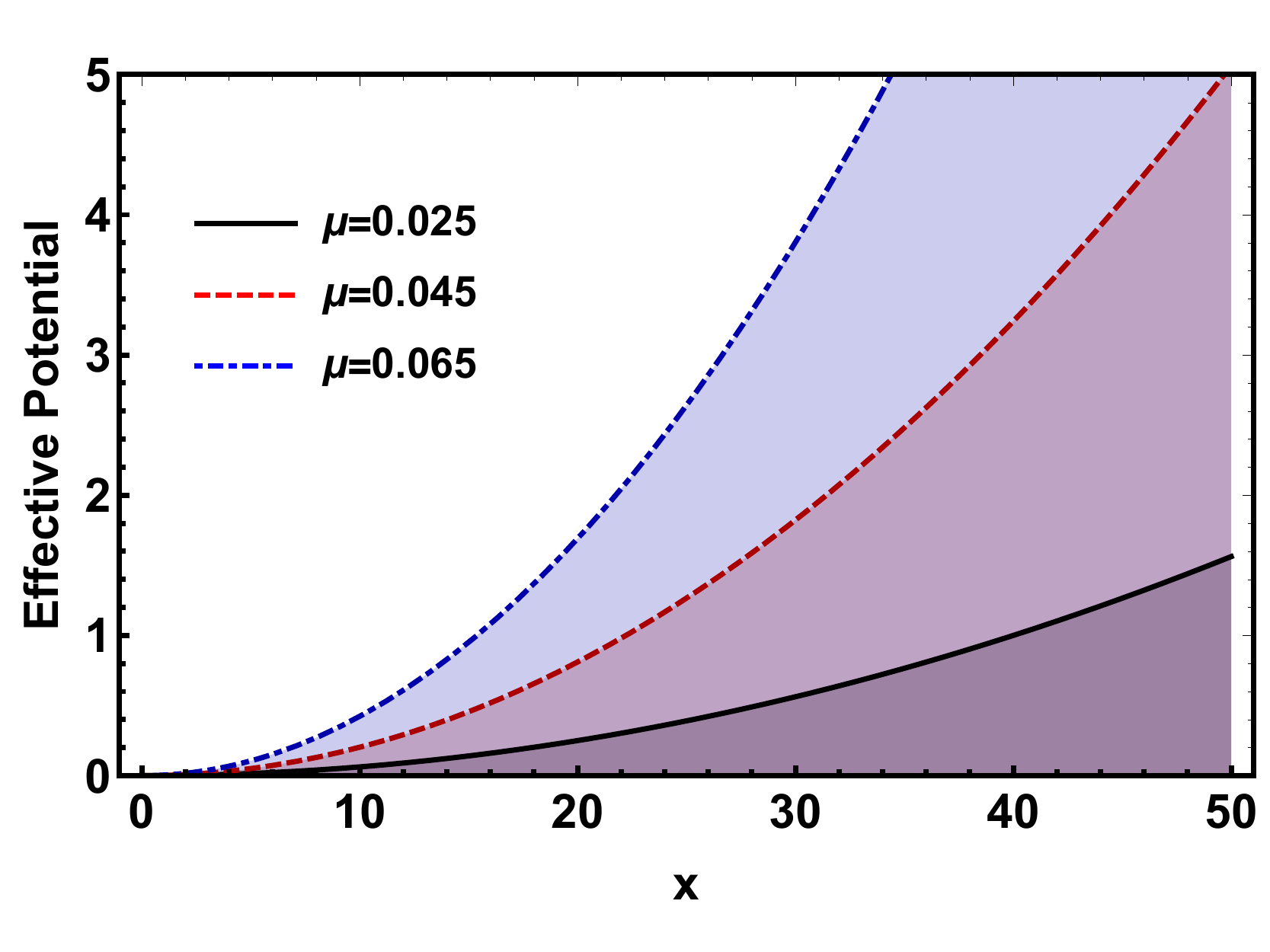}\\
    \vspace{-0.8cm}
    \begin{center} \,  (a) \hspace{7cm} \, \, \, \, (b) \end{center}
    \vspace{-0.8cm}
    \caption{(a) Plot of the $V(x)^2$ produced by the effective mass distribution. (b) Effective potential generated by mass distribution.}
    \label{Lfig2}
\end{figure}

Let us now particularize our study to the case of an effective mass that depends linearly on position, i. e.,
\begin{align}
    m(x)=\mu x \hspace{0.5cm} \text{with} \hspace{0.5cm} x\geq 0.
\end{align}
Here, the $\mu$ parameter adjusts the mass unit, i. e., $\mu$ will have the unit of mass (atomic unit) by length. Fig. \ref{Lfig1} shows the linear profile of the mass. Furthermore, in Fig. \ref{Lfig2} the effective potential (\ref{PoEff}) and the potential (\ref{Potential}) are displayed.

Therefore, to describe the fermions with effective mass, Eq. (\ref{DiracPDM}) is written as
\begin{align}\label{DiracLinearMass}
    -\phi''_1(x)+\mu^2 x^2\phi_1(x)=E^2\phi_1(x).
\end{align}

Investigating the solution of Eq. (\ref{DiracLinearMass}), let us assume the change of variable 
\begin{align}
    \xi=\sqrt{\mu}x \hspace{1cm} \text{with} \hspace{1cm} \xi>0,
\end{align}
which leads us to
\begin{align}\label{p1}
    \phi''_1(\xi)+(K-\xi^2)\phi(\xi)=0,
\end{align}
where $K=E^2/\mu$.

To obtain normalizable wave functions, we must assume
\begin{align}\label{p3}
    \phi_1(\xi)= \text{e}^{-\xi^2/2}h(\xi).
\end{align}
This transformation is useful for obtaining normalizable wave functions. Furthermore, let us point out that some references explicitly suggest the transformation (\ref{p3}) when analytically solving the equations describing quantum oscillators. For more details, see Ref. \cite{Sakurai}.

Adopting the change of variable (\ref{p3}), one obtains
\begin{align}\label{p4}
    h''(\xi)-2\xi h'(\xi)+(K-1)h(\xi)=0.
\end{align}

To investigate the equation (\ref{p4}), we use the Frobenius method. Thus, considering this method, allow us to propose that the solutions of Eq. (\ref{p4}) expressed in terms of the power series are
\begin{align}\label{p5}
    h(\xi)=\sum_{j=0}^{\infty}a_j\xi^j,
\end{align}
which brings us to the recursion relation
\begin{align}\label{p6}
    a_{j+2}=\frac{2j+1-K}{(j+1)(j+2)}a_j.
\end{align}

For the solutions of the expression (\ref{p4}) to be physically acceptable (i. e., the normalizable wave functions), the power series (\ref{p5}) must have finite. So there must be some large $j$ such that the recursion relation (\ref{p6}) generates $a_{n+2}=0$. Furthermore, when $x\to 0$, the wave function is zero due to the mass profile. Thus, assuming these conditions, it is concluded that
\begin{align}
    \label{p7}
    K=2n+1 \hspace{1cm} \text{with} \hspace{1cm} n=1,3,5\dots
\end{align}
i. e.,
\begin{align}
    \label{p8}
    E_n=\sqrt{(2n+1)\mu}.
\end{align}

For allowed values of $K$, the recursion relation takes the form
\begin{align}
    \label{p9}
    a_{j+2}=\frac{-2(n-j)}{(j+1)(j+2)}a_j.
\end{align}

\begin{figure}[ht!]
    \centering
    \includegraphics[height=6.5cm,width=8cm]{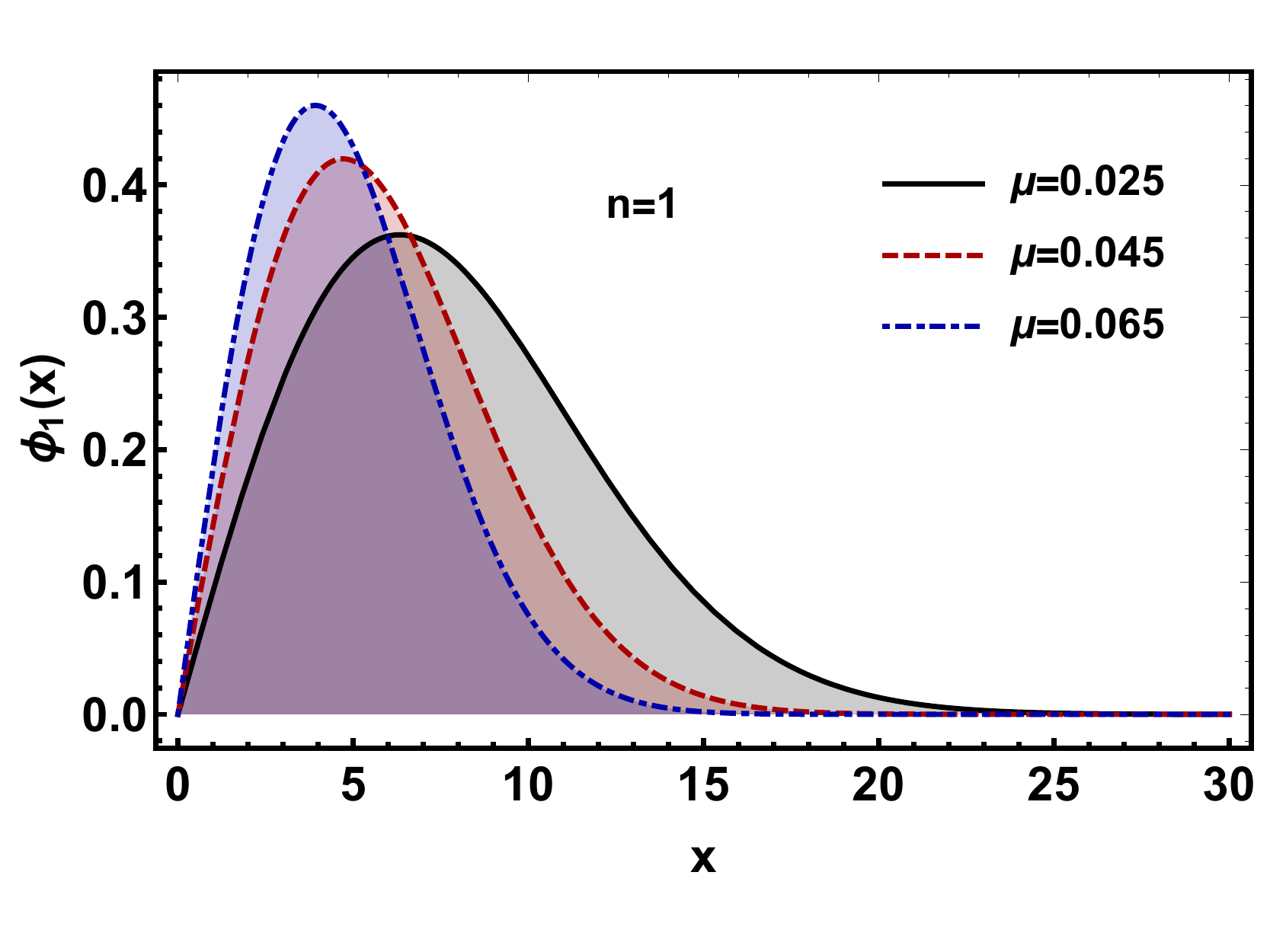}
    \includegraphics[height=6.5cm,width=8cm]{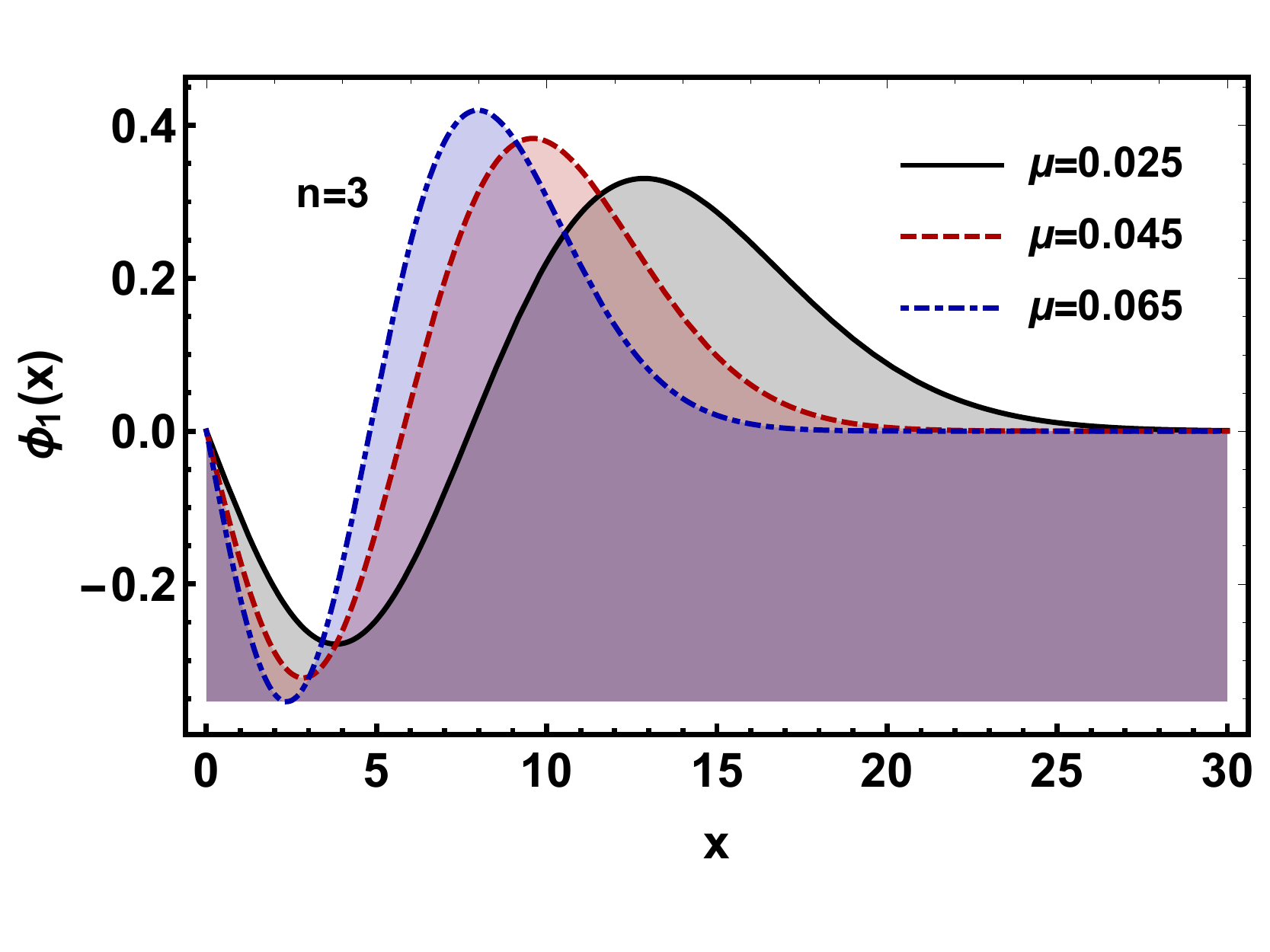}\\
        \vspace{-0.8cm}
    \includegraphics[height=6.5cm,width=8cm]{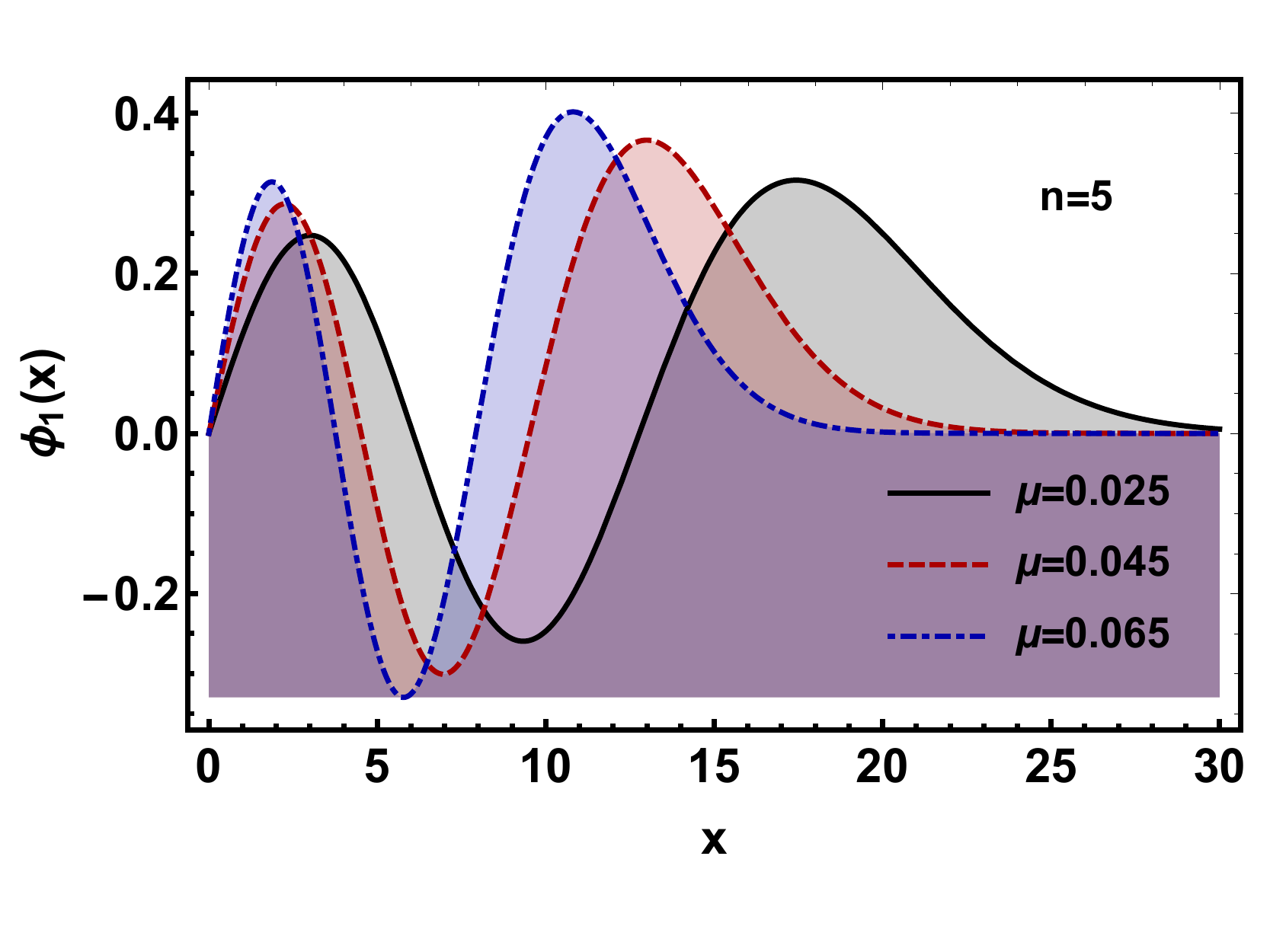}
    \includegraphics[height=6.5cm,width=8cm]{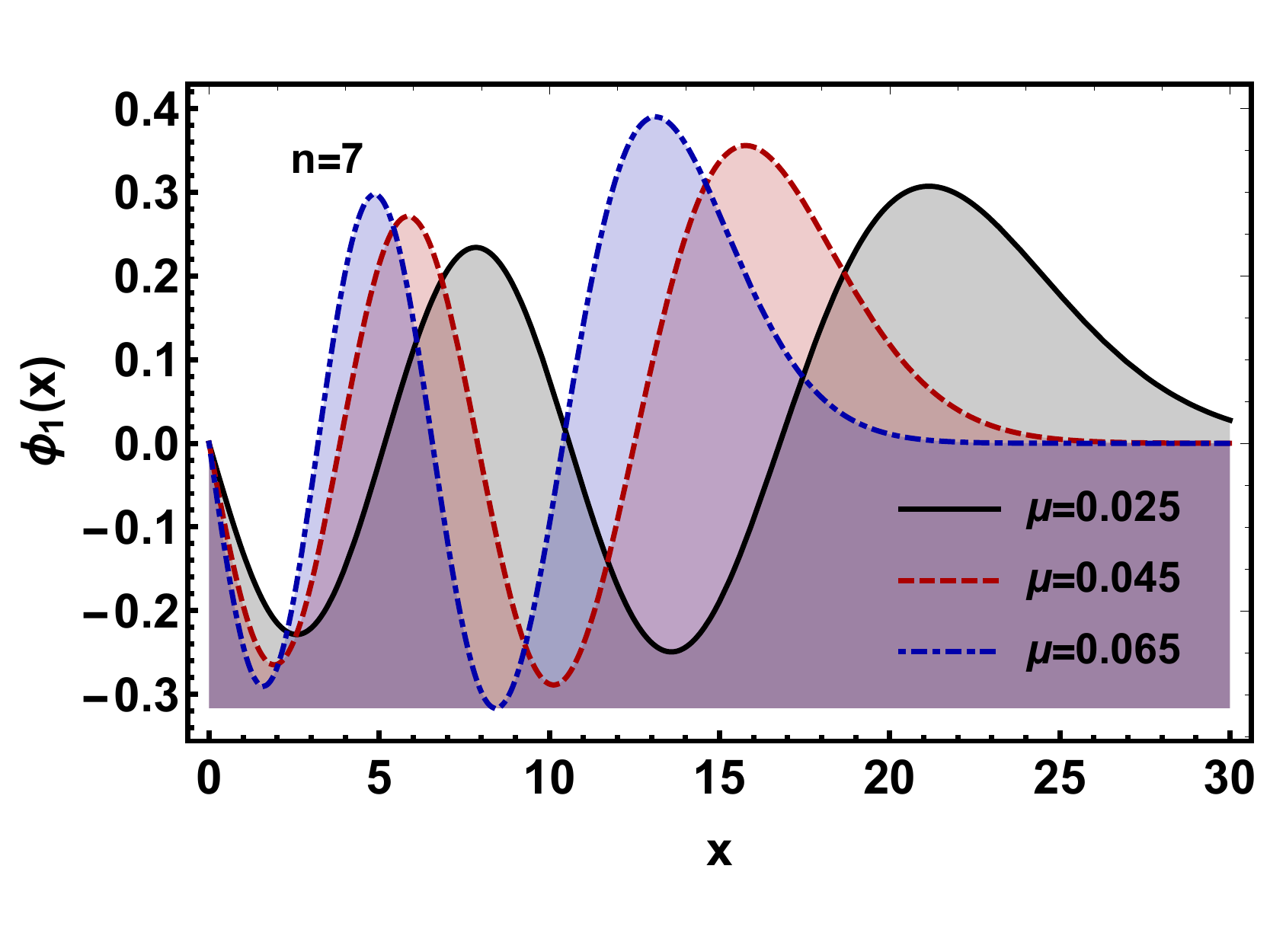}
    \vspace{-0.8cm}
    \caption{Analytical solutions of the first eigenstates of the linear PDM.}
    \label{figLWave}
\end{figure}

Analyzing the recursion relation (\ref{p9}), one concludes that
\begin{align}
    \label{p10}
    h_n(\xi)=A_n H_n(\xi),
\end{align}
where $A_n$ is the normalization constant and $H_n(\xi)$ is the Hermite polynomial, i. e.,
\begin{align}
    \label{Herm}
    H_n(\xi)=(-1)^n\text{e}^{\xi^2}\bigg(\frac{d}{d\xi}\bigg)^n\text{e}^{-\xi^2},
\end{align}
and so, we finally arrive at 
\begin{align}
    \phi_1^{(n)}(x)=A_{n}\text{e}^{-\mu x^2}H_n(\sqrt{\mu}x) \hspace{1cm} \text{with} \hspace{1cm} n=1,3,5,\dots
\end{align}
We show the analytical solutions for the first eigenstates in Fig. \ref{figLWave}.

\subsection{Hyperbolic PDM generating a smooth effective potential barrier}

\begin{figure}[ht!]
    \centering
    \includegraphics[height=6.5cm,width=8cm]{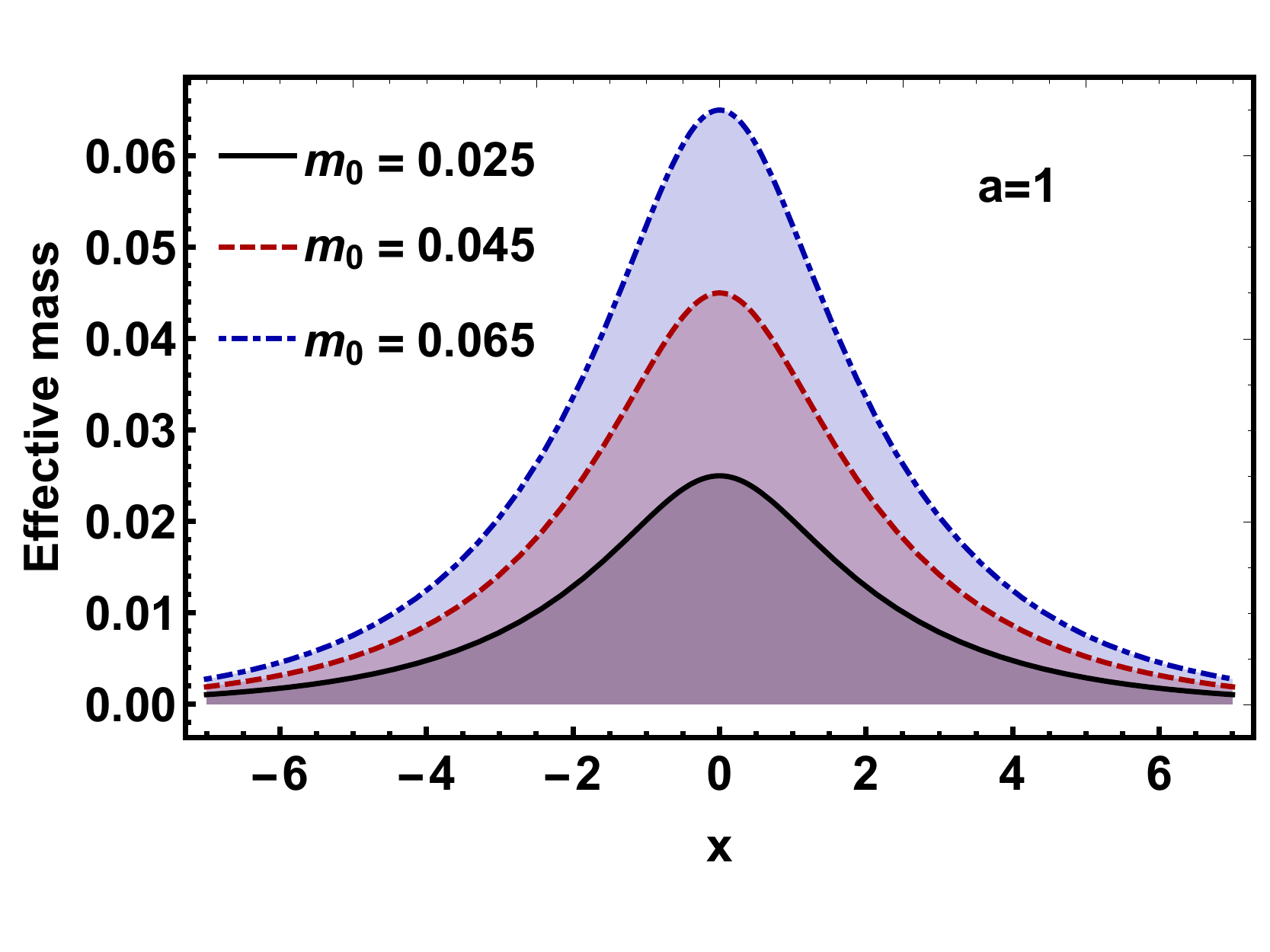}
    \includegraphics[height=6.5cm,width=8cm]{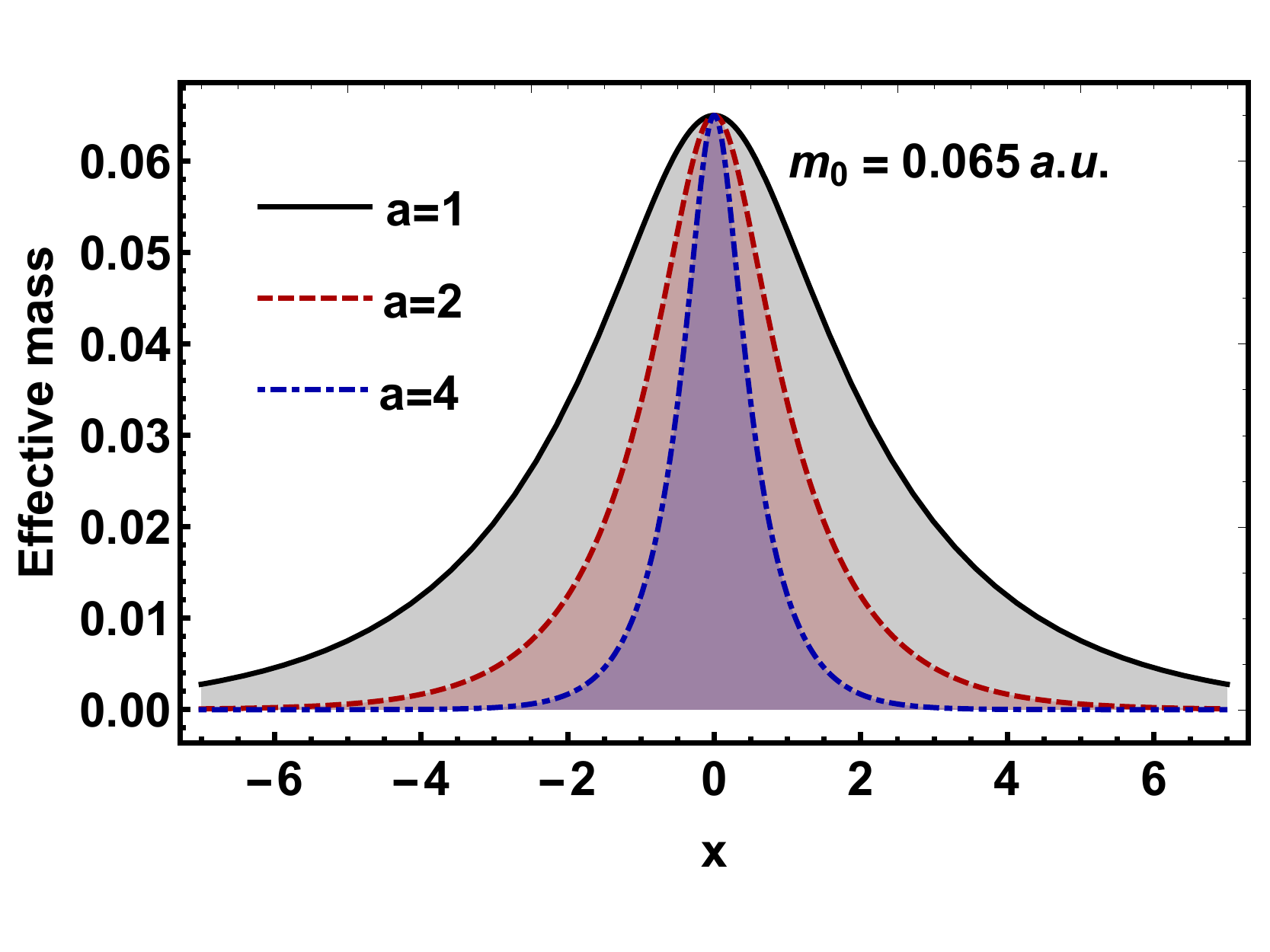}\\
    \vspace{-0.8cm}
    \begin{center} \,  (a) \hspace{7cm} \, \, \, \, (b) \end{center}
    \vspace{-0.8cm}
    \caption{(a) Hyperbolic PDM when $m_0$ varies, and (b) when $a$ varies.}
    \label{fig1}
\end{figure}

\begin{figure}[ht!]
    \centering
    \includegraphics[height=6.5cm,width=8cm]{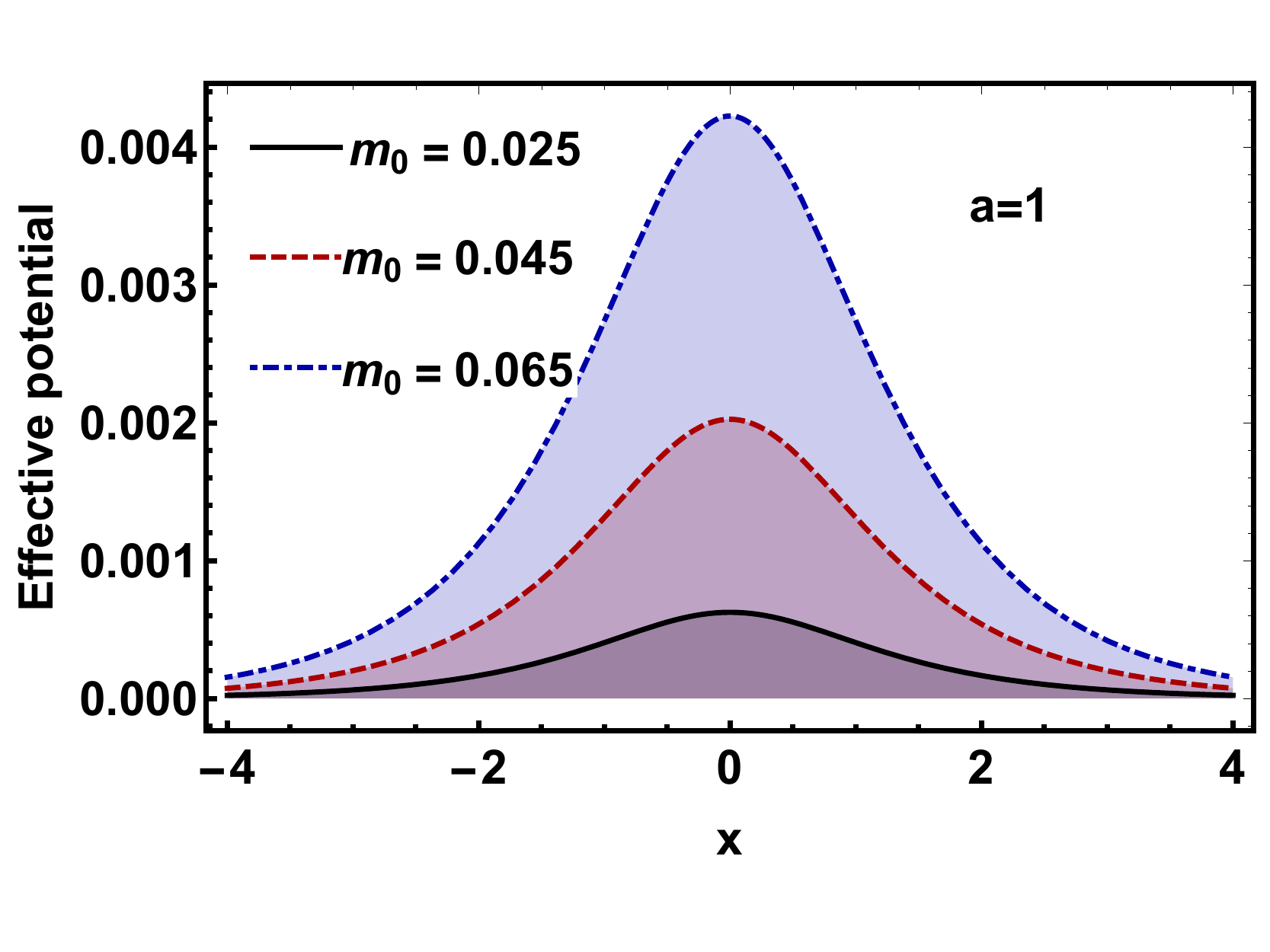}
    \includegraphics[height=6.5cm,width=8cm]{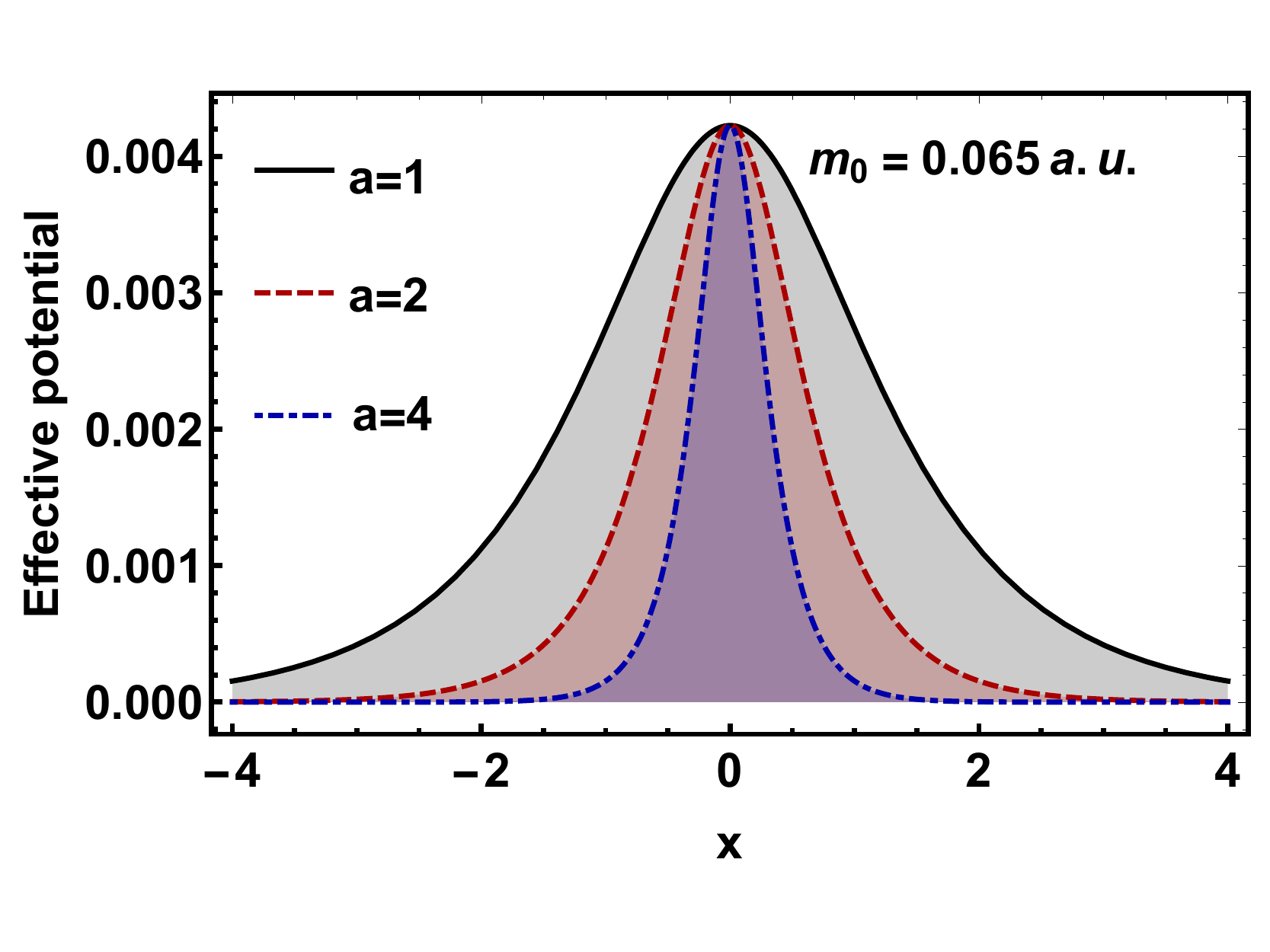}\\
    \vspace{-0.8cm}
    \begin{center} \,  (a) \hspace{7cm} \, \, \, \, (b) \end{center}
    \vspace{-0.8cm}
    \caption{(a) Effective potential when $m_0$ varies, and (b) when $a$ varies.}
    \label{fig2}
\end{figure}

\begin{figure}[ht!]
    \centering
    \includegraphics[height=6.5cm,width=8cm]{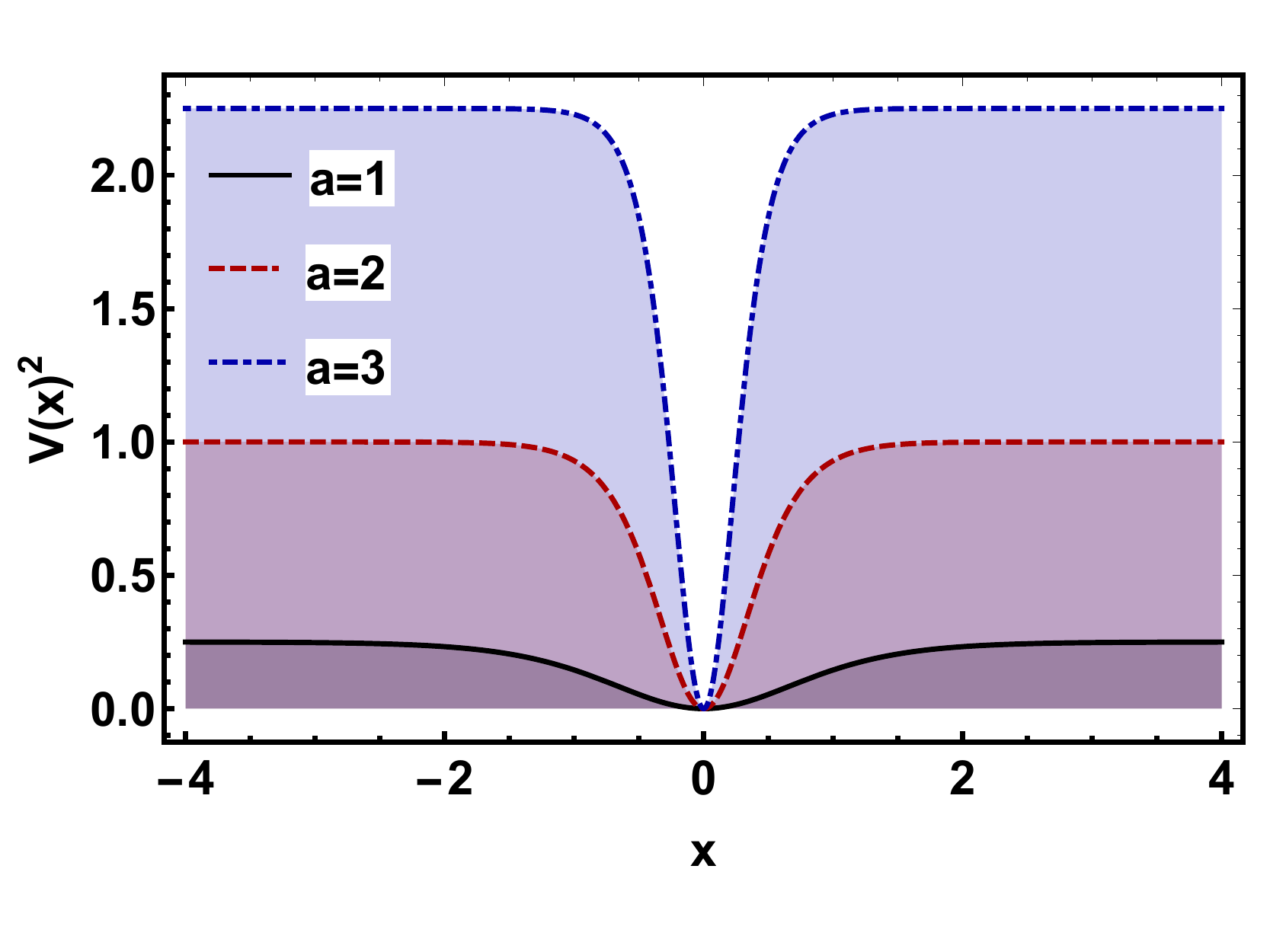}
    \vspace{-0.8cm}
    \caption{Plot of $V(x)^2$ produced by the mass distribution.}
    \label{fig3}
\end{figure}

Now, allow us to particularize the theory for a hyperbolic mass. In this case, the mass profile is
\begin{align}\label{mass}
    m(x)=m_0\,\sqrt{\text{sech}(ax)}.
\end{align}
Here, $m_0$ is the mass distribution. Furthermore, $a$ adjusts the width of the mass distribution. This mass profile (hyperbolic) is known as a soliton. That structure is a topological structure that maintains its shape unchanged when interacting with other solitons and has mass proportional to the $\text{sech}(x)$ \cite{Rajaraman}. Besides, its energy is always finite \cite{Lima1, Rajaraman}. In Fig. \ref{fig1} is exposed the mass profile when the parameters $a$ and $m_0$ are varying. Moreover, we show in Fig. \ref{fig2} the effective potential produced by the mass distribution. In Fig. \ref{fig3}, the profile of the potential squared (\ref{Potential}) is displayed.

For the mass profile (\ref{mass}), Eq. (\ref{DiracPDM}) is rewritten as
\begin{align}
    -\phi''_1(x)+m_{0}^{2}\text{sech}(ax)\phi_1(x)=E^2\phi_1(x).
\end{align}
To solve this equation, let us assume the coordinate change
\begin{align}
    \xi\to\cosh(ax). 
\end{align}
So, the wave function describing $\phi_1(x)$ is rewritten as
\begin{align}\label{HeunPDM}
    \phi''_1(\xi)+\bigg(\frac{1/2}{\xi+1}+\frac{1/2}{\xi-1}\bigg)\phi'_1(\xi)+\bigg[\frac{\mathcal{E}^2\xi-\mu_0^2}{\xi(\xi+1)(\xi-1)}\bigg]\phi_1(\xi)=0, 
\end{align}
with $\mathcal{E}=E/a$ and $\mu_0=m/a$. If we think of our system as a model of solid state physics will be possible to imagine the model as a fermion with an effective mass. In this case, the fermion will compose the crystalline structure with lattice parameter $a$. Furthermore, for this model, the electron will acquire an effective mass that will depend on the position \cite{Kittel,Molinar}. For example, in the case of Al$_{x}$Ga$_{1-x}$As the parameter $m_{0}$ is $0.0665$ a. u. \cite{Adachi}. Therefore, $m_0\ll 1$ is natural.

Allow us to remember that the expression (\ref{HeunPDM}) belongs to the second-order Fuchsian class \cite{Hille}. Besides, it is a particular case of the Heun equation, namely,
\begin{align}\label{Heun}
    H''(y)+\bigg(\frac{\gamma}{y}+\frac{\delta}{y-1}+\frac{\varepsilon}{y-d}\bigg)H'(y)+\frac{\alpha\beta y-q}{y(y-1)(y-d)}H(y)=0.    
\end{align}
The Heun equation parameters must obey the Fuchsian relation, i. e.,
\begin{align}\label{FuchsRel}
    \alpha+\beta+1=\gamma+\delta+\varepsilon.
\end{align}

Moreover, in the neighborhood of each singularity of Heun's equation (\ref{Heun}), two local independent solutions are found. The recurrence relations of the Heun equation are derivative from the Frobenius series \cite{MMaier}. This equation has a set of 192 different expressions of a transformation set of a group of automorphisms \cite{MMaier, Ronveaux}.

We are interested in the solutions of the system (Eq. (\ref{DiracPDM})) in the vicinity of the location of the mass, i. e., $x=0$ (or $\xi=1$). The linearly independent solutions around the singularity $\xi=1$ are
\begin{align}
H^{(1)}(y)=\text{HeunC}(1-d,-q+\alpha\beta, \alpha, \beta, \delta, \gamma; 1-y),
\end{align}
and
\begin{align}\nonumber
    &H^{(2)}(y)=(1-y)^{1-\delta}\times\\
    &\text{HeunC}[1-d, -q+(\delta-1)\gamma d+(\alpha-\delta+1)(\beta-\delta+1), \beta-\delta+1, \alpha-\delta+1, 2-\delta, \gamma; 1-y],
\end{align}
with the characteristic exponents being $0$ and $1-\delta$.

Comparing Heun's equation (\ref{Heun}) with Eq. (\ref{HeunPDM}) and using Fuchsian relation (\ref{FuchsRel}), one obtains
\begin{align}\label{parameters}
    \varepsilon=\frac{1}{2}, \, \, \, \, d=-1, \, \, \, \, \delta=\frac{1}{2}, \, \, \, \, \gamma=0, \, \, \, \, q=\mu_0, \, \, \, \, \alpha=\pm iE \, \, \, \, \text{and} \, \, \, \, \beta=\mp{iE}.
\end{align}

Therefore, the linearly independent solutions of the Eq. (\ref{HeunPDM}) are
\begin{align}\label{solution1}
    \phi^{(1)}_1(\xi)=\text{HeunC}\bigg(2,\, \mathcal{E}^2-\mu_{0}^{2},\, i\mathcal{E},\, -i\mathcal{E},\, \frac{1}{2},\, 0;\, 1-\xi\bigg),
\end{align}
and
\begin{align}\label{solution2}
    \phi^{(2)}_1(\xi)=\sqrt{(1-\xi)}\text{HeunC}\bigg[2,\, \mathcal{E}^2-\mu_{0}^{2}+\frac{1}{4},\, \frac{1}{2} \pm i\mathcal{E},\, \frac{1}{2}\mp i\mathcal{E},\, \frac{3}{2},\, 0;\, 1-\xi\bigg].
\end{align}

In more compact notation, one can write system solutions (\ref{DiracPDM}) as
\begin{align}
    \phi_1(x)=\phi^{(1,2)}_{1}(\cosh(ax)).
\end{align}

\begin{figure}[ht!]
    \centering
    \includegraphics[height=6.5cm,width=8cm]{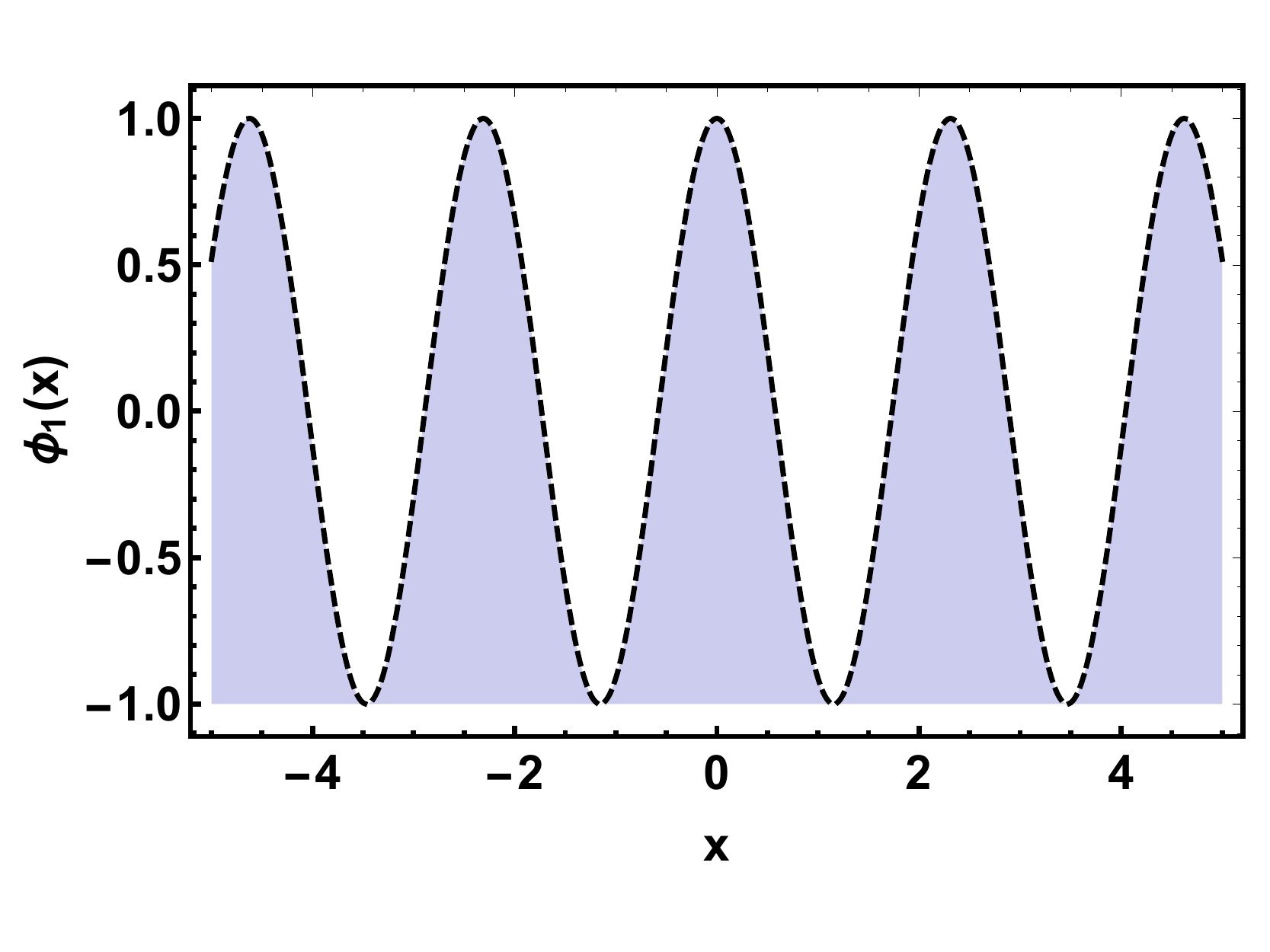}
    \vspace{-0.8cm}
    \caption{Wave eigenfunction $\phi_1(x)$ when $E>0$.}
    \label{fig4}
\end{figure}

Due to the complex profile of the interaction (\ref{Potential}), the wave function will describe a free fermionic particle with energy $E>0$. Physically, this represents a particle that moves freely, i. e., in the absence of interaction (Fig. \ref{fig4}). That is because the particle does not notice any interaction. We show in Fig. \ref{fig4} the eigenfunction of the model.

\section{Final remarks}

In this paper, we perform studies on the one-dimensional relativistic theory for an arbitrary position-dependent mass. First, we considered a fermionic particle with PDM subject to a position-dependent electrostatic interaction. Applying the FW transformation it was possible to analyze the non-relativistic limit. In closing were investigated two mass profiles, i. e., linear PDM and hyperbolic PDM.

Stimulating results arise from investigating an arbitrary mass in the one-dimensional Dirac theory. That results emerge when building a Schr\"{o}dinger-like formalism for the fermion. For example, in seeking a Schr\"{o}dinger-like formalism, it is necessary to assume a complex potential dependent on the spatial distribution of mass. However, in doing so, the system will be a non-Hermitian system. Although the obtained system is non-Hermitian, as the Hamiltonian preserves the $\mathcal{PT}$ symmetry, the energy eigenvalues will be reals. Knowing that the concept of PDM emerges in solid-state physics systems, it is interesting to study a correspondence of our theory with a non-relativistic approach. As a result, one obtains that our model corresponds to the model proposed by Li and Kuhn \cite{LiKuhn}.

When analyzing the linear mass profile it is possible to note the generation of an effective harmonic-like potential that confines the fermion. Furthermore, although the mass has a range of variation in every space, it is perceived that the wave function will only exist for positive values of the position. This result can be seen as a consequence that the particle can never physically admit a negative profile (except for the anti-particle and consequently the $\phi_2$ component). Meantime, it is possible to see that a smooth potential barrier emerges if we consider the hyperbolic mass. However, as the effective potential barrier produced by the particle is weak compared to the particle energy, the solution of the system will be a flat wave. This result is interesting because, in a non-relativistic theory, this mass profile will produce effective potentials with the ability to confine it, as discussed in Ref. \cite{Lima1}.

\section*{Acknowledgments}
\hspace{0.5cm} The authors thank the Conselho Nacional de Desenvolvimento Cient\'{\i}fico e Tecnol\'{o}gico (CNPq), n$\textsuperscript{\underline{\scriptsize o}}$ 309553/2021-0 (CASA), and Coordena\c{c}\~{a}o de Aperfei\c{c}oamento do Pessoal de N\'{i}vel Superior (CAPES), for financial support.

\end{document}